\documentstyle[12pt]{article}

\newcommand{\ddouble}{{\partial^{^{\kern-6pt \leftrightarrow}}}}

\newcommand{\dfrac}[2]{{\displaystyle{\frac{#1}{#2}}}}

\newcommand{\dint}{\displaystyle{\int}}
\newcommand{\equ}[1]{(\ref{#1})}
\newcommand{\es}{\\[3mm]}
\newcommand{\beq}{\begin{equation}}
\newcommand{\eqn}[1]{\label{#1}\end{equation}}
\newcommand{\ba}{\begin{array}}
\newcommand{\ea}{\end{array}}

\def\vf{\varphi}

\def\l{\lambda}
\def\m{\mu}
\def\n{\nu}

\def\F{\Phi}
\def\G{\Gamma}
\def\psi{\Psi}

\def\S{\Sigma}

\def\cf{{\cal F}}

\def\inbar{\vrule height1.5ex width.4pt depth0pt}
\def\rlx{\relax\leavevmode}
\def\I{\leavevmode\hbox{\small1\kern-3.8pt\normalsize1}}
\def\openone{\leavevmode\hbox{\small1\kern-3.3pt\normalsize1}}
\def\Ione{\rlx{\rm 1\kern-2.7pt l}}
\def\Ik{\rlx{\rm I\kern-.18em k}}
\def\IC{\rlx\leavevmode
             \ifmmode\mathchoice
                    {\hbox{\kern.33em\inbar\kern-.3em{\rm C}}}
                    {\hbox{\kern.33em\inbar\kern-.3em{\rm C}}}
                    {\hbox{\kern.28em\sinbar\kern-.25em{\rm C}}}
                    {\hbox{\kern.25em\ssinbar\kern-.22em{\rm C}}}
             \else{\hbox{\kern.3em\inbar\kern-.3em{\rm C}}}\fi}
\def\IP{\rlx{\rm I\kern-.18em P}}
\def\IR{\rlx{\rm I\kern-.18em R}}
\def\IN{\rlx{\rm I\kern-.20em N}}

\def\llsymbol#1{\@llsymbol{\@nameuse{c@#1}}}
\def\@llsymbol#1{\ifcase#1\or {}\or {'}\or {''}\or {'''}\or
   {''''}\or {'''''}\or  \else\@ctrerr\fi\relax}
\newcounter{contador}

\newcommand{\ol}\overline
\newcommand{\ti}\tilde
\newcommand{\wt}\widetilde
\newcommand{\wh}\widehat
\newcommand{\bv}\breve
\newcommand{\dg}\dagger

\newcommand{\C}{^{\mbox{\scriptsize c}}}

\newcommand{\T}{^{\mbox{\scriptsize T}}}

\newcommand{\be}{\begin{equation}}
\newcommand{\ee}{\end{equation}}
\newcommand{\bl}{\begin{eqnarray}&}
\newcommand{\el}{&\end{eqnarray}}
\newcommand{\bq}{\begin{eqnarray}}
\newcommand{\eq}{\end{eqnarray}}

\newcommand{\pa}{\partial}

\renewcommand{\theequation}{\thesection.\arabic{equation}}

\begin{document}

{\hfill\parbox{50mm}{\large hep-th/9711191\\
                             CBPF-NF-052/97\\
                             UFES--DF--OP97/2}} \vspace{3mm}

\begin{center}
{{\LARGE {\bf On the Non-Renormalization Properties\\[3mm]of Gauge Theories 
\\[4mm]with a Chern-Simons Term}}} \vspace{7mm}

{\large Oswaldo M. Del Cima$^{{\rm (a),(b),}}$\footnote{Current 
address: 
{\it Institut f\"ur Theoretische Physik, 
Technische Universit\"at Wien, Wiedner Hauptstrasse 8-10 - 
A-1040 - Vienna - Austria}.}, 
Daniel H.T. Franco$^{{\rm (b),}}$\footnote{Supported by 
the {\it Conselho Nacional de 
Desenvolvimento 
Cient\'\i fico e Tecnol\'ogico (CNPq)}.},\\[2mm]
Jos\'e A. Helay\"el-Neto$^{{\rm 
(b)}}$ 
and Olivier Piguet$^{{\rm (c),2,}}$\footnote{Supported in part by the 
{\it Swiss 
National Science Foundation}.}${}^{,}$\footnote{On leave of 
absence from {\it D\'epartement de Physique Th\'eorique, 
Universit\'e de Gen\`eve, 24 quai E. Ansermet - CH-1211 - 
Gen\`eve 4 - Switzerland}.} } 
\vspace{4mm}

$^{{\rm (a)}}$ {\it Pontif\'\i cia Universidade Cat\'olica do 
Rio de Janeiro
(PUC-RIO), \\Departamento de F\'\i sica, \\Rua Marqu\^es de S\~ao 
Vicente, 
225 - 22453-900 - Rio de Janeiro - RJ - Brazil.}

$^{{\rm (b)}}$ {\it Centro Brasileiro de Pesquisas 
F\'\i sicas (CBPF), 
\\Departamento de
Teoria de Campos e Part\'\i culas (DCP),\\Rua Dr. Xavier 
Sigaud 150 - 
22290-180 - Rio de Janeiro - RJ - Brazil.}

$^{{\rm (c)}}$ {\it Universidade Federal do Esp\'\i rito 
Santo (UFES), 
\\CCE,
Departamento de F\'\i sica, \\Campus Universit\'ario de Goiabeiras - 
29060-900 - Vit\'oria - ES - Brazil.}
\vspace{4mm} 

{\it E-mails: delcima@tph73.tuwien.ac.at, 
dfranco@cbpfsu1.cat.cbpf.br, 
helayel@cbpfsu1.cat.cbpf.br, piguet@cce.ufes.br.} 
\end{center}

\newpage

\begin{abstract}
Considering three-dimensional Chern-Simons theory, either coupled to matter
or with a Yang-Mills term, we show the validity of a trace identity, playing
the role of a local form of the Callan-Symanzik equation, in all orders of
perturbation theory. From this we deduce the vanishing of the 
$\beta$-function associated to the Chern-Simons coupling constant and 
the full finiteness in the case of the Yang-Mills Chern-Simons theory. 
The main ingredient in the proof of the latter property is the 
noninvariance of the Chern-Simons form under the gauge transformations. 
Our results hold for the three-dimensional Chern-Simons model in a general 
Riemannian manifold.
\end{abstract}



\section{Introduction}

Topological field theories (see~\cite{bbr-pr91} for a general review and
references) are a class of gauge model interesting from a physical point of
view. In particular, their observables are of topological nature. An
important topological model that has received much attention in the last
years is the Chern-Simons model in three 
dimensions~\cite{s-btc,witten,wcmp121}. 
In contrast to the usual Yang-Mills gauge theories,
Chern- Simons theories, which include the case of three-dimensional 
gravity~\cite{witten-grav,deser-grav},
have some remarkable features. 
Besides their relevance in connection with
the possibility of getting nonperturbative results more easily and their
relation with the two-dimensional conformal field theories~\cite{moore},
they show very interesting perturbative 
features  such as ultraviolet 
finiteness~\cite{witten-grav,deser-grav,guadagnini}. 
 A complete and rigorous proof of the latter property has been 
given in~\cite{blasi} for $D$ $=$ $1+2$ Chern-Simons theories in
the Landau gauge. In the more
general case where these models are coupled to matter fields, the
Chern-Simons coupling constant keeps unrenormalized, the corresponding 
$\beta $-function remaining vanishing. This has been rigorously proved 
in~\cite{maggiore} for the Abelian Chern-Simons theory coupled to 
scalar matter
fields. For the nonabelian theory coupled to spinorial
matter fields, an argument based on the assumed existence of an invariant
regularization has been proposed by the authors of~\cite{pronin}.

On the other hand, there exist studies in the literature concerning 
the Yang-Mills Chern-Simons (YMCS) theory~\cite{deser,martin,pisa,ruiz}. In
particular the finiteness at one loop was detected in 
refs.~\cite{deser,pisa}, 
where explicit expressions for one-loop radiative corrections can be
found. Later on in~\cite{ruiz} has been showed that no UV divergences arise
up to two-loops in perturbation theory, combining this result with
finiteness by power counting at higher loops to argument that theory is
finite at any perturbative order. Recently, similar arguments has been used
to prove the finiteness for the $N=1$ supersymmetric version of the YMCS
theory~\cite{RN}. A cohomological study of $N=2$ YMCS theory in a
nonsupersymmetric gauge has been performed in~\cite{mpr}, where a partial
proof of the finiteness is given. 

  Two recent papers~\cite{silvio,silvio2} have
shown the equivalence of the YMCS theory with a 
pure Cherns-Simons theory, in the classical approximation.
One of our results, namely the complete ultraviolet
finiteness of the YMCS, although proved in perturbation theory only,
indicates that this equivalence could hold for the full quantum theory.
Let us  finally mention that Ref.~\cite{silvio2}, 
appeared after completion of the present paper, beyond of
expanding the classical results of~\cite{silvio} and arguing in favour of 
a possible quantum generalization, shows that, 
at the quantum level,
the YMCS theory is finite up to possible field amplitude renormalizations. 

Our first purpose in this paper is to give a rigorous proof of the vanishing
of the Chern-Simons $\beta$-function in the presence of general scalar and
spinorial matter, introduced in the most general way compatible with
renormalizability. We shall avoid the necessity of invoking a particular
regularization procedure, by using the ``algebraic'' method of
renormalization~\cite{pigsor}, which only relies on general theorems of
renormalization theory. We think indeed that, due to the presence of the
antisymmetric Levi-Civita tensor, it is difficult to establish an invariant
regularization without encountering problems at some or other stage of the
argument. Thus, to the contrary of the authors of~\cite{pronin}, we shall
neither assume an invariant regularization nor the explicit ``multiplicative
renormalizability'' at the level of the Lagrangian which would follow from
this assumption.

Our  first result generalizes 
that of~\cite{maggiore}, the gauge group being now an
arbitrary compact Lie group. We exploit the scaling properties of the
theory, described by the Callan-Symanzik equation, which is the anomalous
Ward identity for scale invariance. More precisely, as in~\cite{maggiore},
we make use of a local form of the Callan-Symanzik equation. But, instead of
introducing an external dilatation field as in~\cite{maggiore}, beyond the
external metric or dreibein field, we only consider the latter, considered
as a source for the BRS invariant energy-momentum tensor. The conservation
of the energy-momentum tensor is expressed through the Ward identity which
characterizes the invariance of the theory under the diffeomorphisms. Our
``local Callan-Symanzik equation'' then is the anomalous Ward identity for
the BRS invariant trace of the energy-momentum tensor, the so-called ``trace
identity''. As expected~\cite{maggiore,pronin}, the nonrenormalization of
the Chern-Simons coupling, which amounts to the vanishing of the
corresponding $\beta$-function, is traced back to the fact that the
integrand of the Chern-Simons action is not BRS invariant but transforms as
a total derivative.

Of course the $\beta$-functions associated with the other couplings -- the
self-couplings of the matter fields -- do not vanish in general.

 Our second result is a purely algebraic proof of the all-order,
 complete 
finiteness of the YMCS theory, using the same techniques, but exploiting the
superrenormalizability of the model. To our knowledge, up to now, no such
algebraic proof had been given. 
  Let us note that, although the
proof of finiteness given in~\cite{silvio2} confirms our 
second result 
in quite an independent way, there are differences which should be
noted. First, the result of~\cite{silvio2} is more general than ours
since it holds for a more general set of theories, which are
not necessarely power-counting renormalizable. Second, this result holds
only up to possible field amplitude renormalizations.
However, restricting
ourselves to the power-counting renormalizable case, we are also able to
show the complete finiteness, including the nonrenormalization of the
field amplitudes, as it is the case for the pure Chern-Simons 
theory~\cite{guadagnini,blasi}. 

Since we are working with an external dreibein which is not necessary flat,
our results hold for a curved manifold, as long as its topology remains that
of flat ${\cal R}^3$, with asymptotically vanishing curvature. It is the
latter two restrictions which allow us to use the general results of
renormalization theory, established in flat space. Indeed, we may then
expand in the powers of the difference between the curved and flat space
dreibeins, this difference being considered as an external field in flat
space, fastly decreasing asymptotically.

The paper is organized as follows. The Chern-Simons theory coupled to
matter, in a curved three-dimensional Riemannian manifold described in terms
of external dreibein and spin connection fields, is introduced in 
Section~\ref{sect2}, together with its symmetries. The renormalizability 
of the model is sketched in Section~\ref{sect3}, and our argument 
leading to the
vanishing of the Chern-Simons $\beta $-function is presented in 
Section~\ref{sect4}. The application of our techniques to the 
finiteness of the YMCS
theory is presented in section~\ref{sect5}, followed by our conclusions. The
paper is completed with two Appendices: In the App. A we compute the
classical trace identity whose quantum extension leads to the local form of
the Callan-Symanzik equation. The App. B recalls some properties of the
short distance Wilson expansion which are used in the main text.


\section{The Chern-Simons Model Coupled With Matter in Curved Space-Time}

\label{sect2}

\setcounter{equation}{0} 

\subsubsection*{The Classical Action}

The gauge field $A_\mu ^a(x)$ lies in the adjoint representation of the
gauge group $G$, a general compact Lie group with Lie algebra 
\[
\left[ X_a,X_b\right] =if_{ab}{}^cX_c\,\ . 
\]
The scalar matter fields $\varphi _i(x)$ and the spinor matter fields 
$\Psi_A(x)$ are in some representations of $G$, the generators being
represented by the matrices $T_a^{(\varphi )}$ and $T_a^{(\Psi )}$,
respectively.

Space-time is a three-dimensional Riemannian manifold ${\cal M}$, with
coordinates $x^\mu $, $\mu =0,1,2$. It is described by a dreibein field 
$e_\mu ^m(x)$ and its inverse $e_m^\mu (x)$, $\mu $ being a world index and 
$m$ a tangent space index. The spin connection $\omega _\mu^{mn}(x)$ is
not an independent field, but depend on the dreibein due to the vanishing
torsion condition. The metric tensor and its inverse read 
\begin{equation}
g_{\mu \nu }\left( x\right) =\eta _{mn}\,e_\mu ^m\left( x\right) e_\nu
^n\left( x\right) \,\,,\quad g^{\mu \nu }\left( x\right) =\eta
^{mn}\,e_m^\mu \left( x\right) e_n^\nu \left( x\right) \,\,,
\end{equation}
$\eta _{mn}$ being the tangent space flat metric. We denote by \thinspace $e$
\thinspace the determinant of $e_\mu ^m$.

As explained in the Introduction, we assume the manifold ${\cal M}$ to be
topologically equivalent to ${\cal R}^3$ and asymptotically flat.

\noindent The invariances of the theory are:

1) {\it Gauge invariance}. The infinitesimal gauge tranformations read --
anticipating, we write them as BRS transformations, i.e. we replace their
infinitesimal parameters by the anticommuting Faddeev-Popov ghost fields 
$c^a(x)$; we also introduce the antighost fields ${\bar c}^a(x)$ and the
Lagrange multiplyier fields $b^a(x)$ which will be used later on in order to
define the gauge fixing condition -- 
\begin{equation}
\begin{array}{ll}
sA_\mu ^a=-D_\mu c^a
 \equiv -\left( \pa_\m c^a + f_{bc}{}^a A_\m^b c^c \right)\,,
\qquad & sc^a=\frac 12 f_{bc}{}^a c^b c^c\,, \\[3mm] 
s\Psi _{_A}=\,i\,c^aT_a^{(\Psi )}{}_A{}^B\Psi_B\,\,,\qquad & s{\bar{c}}
^a=b^a\,, \\[3mm] 
s\varphi _i=\,i\,c^aT_a^{(\varphi )}{}_i{}^j\varphi_j\,\,,\qquad & sb^a=0\,.
\end{array}
\label{BRSnonab1}
\end{equation}
The variation of the ghost $c$ is chosen such as to make the BRS operator $s$
nilpotent: 
\[
s^2 = 0\, . 
\]

2) {\it Invariance under diffeomorphisms}. The infinitesimal diffeomorphisms
read 
\begin{equation}
\begin{array}{l}
\delta _{{\rm diff}}^{(\varepsilon )}F_\mu ={\cal L}_\varepsilon F_\mu
=\varepsilon ^\lambda \partial _\lambda F_\mu +\left( \partial _\mu
\varepsilon ^\lambda \right) F_\lambda \,,\quad F_\mu =A_\mu ^a,\ e_\mu ^m\,,
\\[3mm] 
\delta _{{\rm diff}}^{(\varepsilon )}\Phi ={\cal L}_\varepsilon \Phi
=\varepsilon ^\lambda \partial _\lambda \Phi \,,\quad \Phi =\varphi _i,\
\Psi _A,\ b^a,\ c^a,\ {\bar{c}}^a\,,
\end{array}
\label{diff-fields}
\end{equation}
where ${\cal L}_\varepsilon $ is the Lie derivative along the vector field 
$\varepsilon ^{\mu}(x)$ -- the infinitesimal parameter of the transformation.

3) {\it Local Lorentz invariance}. The infinitesimal transformations, with
infinitesimal parameters $\lambda_{[mn]}$, are given by 
\begin{equation}
\delta_{{\rm Lorentz}}^{(\lambda)} \Phi = \frac12 \lambda_{mn}\Omega^{mn}
\Phi\,,\quad \Phi = \mbox{any field}\,,  \label{lor-fields}
\end{equation}
with $\Omega^{[mn]}$ acting on $\Phi$ as an infinitesimal Lorentz matrix in
the appropriate representation.

The most general classical, power-counting renormalizable action invariant
under the diffeomorphisms and local Lorentz transformations, and gauge
invariant -- i.e. BRS-invariant and independent of the ghost fields -- is of
the form 
\begin{equation}
\begin{array}{l}
\Sigma _{{\rm inv}}=\displaystyle{\int }d^3x\,\left( \kappa \,\varepsilon
^{\mu \nu \rho }\left( A_\mu ^a\partial _\nu A_\rho ^a+\frac 13f_{abc}A_\mu
^aA_\nu ^bA_\rho ^c\right) \right. \\[3mm] 
\left. \phantom{\S_{\rm inv} = \dint d^3x\,(}+e\,\left( \,i\,\bar{\Psi}
_{_A}\gamma ^\mu {\cal D}_\mu \Psi _{_A}+\frac 12{\cal D}_\mu \varphi _i
{\cal D}^\mu \varphi _i-{\cal V}(\varphi ,\,\Psi )\right) \right) \,,
\end{array}
\label{g-inv-action1}
\end{equation}
with the generalized covariant derivative defined by 
\begin{equation}
{\cal D}_\mu \Phi (x)\equiv \left( \partial _\mu -i\,{A_\mu ^a}(x)T_a^{(\Phi
)}+\frac 12\omega _\mu ^{mn}(x)\Omega _{mn}\right) \Phi (x)\,\,.
\end{equation}
The function ${\cal V}(\varphi ,\,\Psi )$ defines the selfinteractions of
the matter fields and their masses: 
\begin{equation}
{\cal V}\left( \varphi ,\,\Psi \right) =\frac 16\lambda \,\varphi ^6+\frac
12y\,\bar{\Psi}\Psi \varphi ^2+\mbox{mass terms}+
\mbox{dimensionful
couplings}\,,  \label{potential}
\end{equation}
whith the short-hand notation 
\[
\begin{array}{l}
\lambda \,\varphi ^6=\lambda _{ijklmn}\varphi _i\varphi _j\varphi _k\varphi
_l\varphi _m\varphi _n\,\,, \\[3mm] 
y\,\bar{\Psi}\Psi \varphi ^2=y_{ABmn}\bar{\Psi}_A\Psi _B\varphi _m\varphi
_n\,\,,
\end{array}
\]
$\lambda $ and $y$ being invariant tensors of the gauge group. We do not
write explicitly the mass terms and the dimensionful couplings since we are
interested in the scaling properties in the high energy-momentum limit. 

\subsubsection*{Gauge Fixing}

The gauge fixing is of the Landau type. It is implemented by adding to the
gauge invariant action (\ref{g-inv-action1}) the term 
\begin{equation}
\Sigma_{{\rm gf}} = -\,s\,\displaystyle{\int} d^3x\,\,e\,g^{\mu\nu}\partial_
\mu \bar{c}_a A_\nu^a \,=\, -\, \displaystyle{\int} d^3x\,\,e\,g^{\mu\nu}
\left( \partial_\mu b_a A_\nu^a + \partial_\mu \bar{c}_a D_\nu c^a \right)
\,,  \label{g-fix-action}
\end{equation}
which is BRS invariant due to the nilpotency of $s$.

Moreover, because of the nonlinearity of some of the BRS transformations 
(\ref{BRSnonab1}), we have also to add a term giving their coupling with
external fields, the ``antifields'' $A_a^{*\mu }$, $c_a^{*}$, $\Psi _A^{*}$, 
$\varphi _i^{*}$: 
\begin{equation}
\Sigma _{{\rm ext}}=\displaystyle{\int }d^3x\sum_{\Phi =A_\mu
^a,\,c^a,\,\Psi _A,\,\varphi _i}\Phi ^{*}s\Phi \,.  \label{ext-action}
\end{equation}
The antifields are tensorial densities: they transform under the
diffeomorphisms as 
\[
\begin{array}{l}
\delta _{{\rm diff}}^{(\varepsilon )}A_a^{*\mu }={\cal L}_\varepsilon
A_a^{*\mu }=\partial _\lambda \left( \varepsilon ^\lambda A_a^{*\mu }\right)
-\left( \partial _\lambda \varepsilon ^\mu \right) A_a^{*\lambda } \\[3mm] 
\delta _{{\rm diff}}^{(\varepsilon )}\Phi ^{*}={\cal L}_\varepsilon \Phi
^{*}=\partial _\lambda \left( \varepsilon ^\lambda \Phi ^{*}\right) \,,\quad
\Phi ^{*}=c_a^{*},\,\varphi _i^{*},\,\Psi _A^{*}\,.
\end{array}
\]
Their transformations under local Lorentz symmetry are obvious. They are
moreover BRS invariant.

  From now on we consider the total action 
\begin{equation}
\Sigma = \Sigma_{{\rm inv}} + \Sigma_{{\rm gf}}+ \Sigma_{{\rm ext}}\,.
\label{total-action}
\end{equation}

\subsubsection*{The Functional Identities}

The various symmetries of the model as well as the gauge fixing we have
choosen may be expressed as functional identities obeyed by the classical
action (\ref{total-action}).

\noindent The BRS invariance is expressed through the Slavnov-Taylor\
identity 
\begin{equation}
{\cal S}(\Sigma )\,\, =\,\, \displaystyle{\int} d^3x\, \displaystyle{
\sum_{\Phi =A_\mu^a,\,c^a,\,\Psi_{_A},\,\varphi_i}^{}} {\displaystyle{\frac{
\delta\Sigma }{\delta \Phi^{*}}}}{\displaystyle{\frac{\delta \Sigma }{\delta
\Phi }}} \, +b\, \Sigma = 0 \,,\quad\mbox{with } \ b = \displaystyle{\int}
d^3x\, b^a{{\displaystyle{\frac{\delta}{\delta \bar{c}^a}}}}\,.
\label{slavnonab1}
\end{equation}
For later use we introduce the linearized Slavnov-Taylor\ operator 
\begin{equation}
{\cal B}_{\Sigma }\,\,=\,\,\displaystyle{\int} {d}^3x\, \displaystyle{\
\sum_{\Phi =A_\mu^a,\,c^a,\,\Psi_{_A},\,\varphi_i}^{}} \left( {\displaystyle{
\ \frac{\delta\Sigma }{\delta \Phi^{*}}}}{\displaystyle{\frac{\delta}{\delta
\Phi }}} + {\displaystyle{\frac{\delta\Sigma }{\delta \Phi}}}{\displaystyle{
\ \frac{\delta}{\delta \Phi^*}}}\right) \, + b\,.  \label{linear}
\end{equation}
${\cal S}$ and ${\cal B}$ obey the algebraic identity 
\begin{equation}
{\cal B}_\cf {\cal B}_\cf {\cal F}^{\prime}+ \left({\cal B}_{{\cal F}
^{\prime}}-b\right) {\cal S}({\cal F}) = 0\,,  \label{fcondnonab1}
\end{equation}
${\cal F}$ and ${\cal F}^{\prime}$ denoting arbitrary functionals of ghost
number zero. From the latter follow 
\begin{equation}
{\cal B}_{{\cal F}}\,{\cal S}({\cal F})=0\;\;,\;\;\;\forall \;{\cal F}\;\;\;,
\label{nilpot1}
\end{equation}
\begin{equation}
\left({\cal B}_{{\cal F}}\right)^2 = 0\;\;\;{\mbox{if}}\;\;\; {\cal S}({\cal 
F})=0\,.  \label{nilpot3}
\end{equation}
In particular, since the action $\Sigma$ obeys the Slavnov-Taylor\ identity,
we have the nilpotency property (\ref{slavnonab1}): 
\begin{equation}
\left({\cal B}_{\Sigma }\right)^2=0\,\,.  \label{nilpot2}
\end{equation}
In addition to the Slavnov-Taylor identity (\ref{slavnonab1}), the action 
(\ref{total-action}) satisfies the following constraints:

\noindent -- the Landau gauge condition: 
\begin{equation}
{\displaystyle{\frac{\delta \Sigma}{\delta b_a}}}= \partial _\mu \left(
eg^{\mu \nu }A_\nu^a\right) \,\, ,  \label{landau1}
\end{equation}

\noindent -- the ``antighost equation'', peculiar to the 
Landau gauge~\cite{bl-pig-sor}, 
\begin{equation}
{\bar{{\cal G}}}^a\Sigma \,\,=\,\, 
\displaystyle{\int} d^3x
 \left({\displaystyle{\ \frac{\delta}{\delta c^a}}}
 +f^{\,abc}\bar{c}_b {\displaystyle{\frac{\delta}{\delta b^c}}}\right)
   \Sigma\,\,=\,\, 
\Delta_{{\rm cl}}^a\,,  \label{antighost}
\end{equation}
with 
\[
\Delta_{{\rm cl}}^a=\displaystyle{\int}
d^3x \left(f^{\,abc}\left( A_b^{*\mu }A_{c\mu } - c_b^{*}c_c\right)
+ i{\bar\Psi}^* T_a^{(\Psi)}\Psi - i\varphi^* T_a^{(\varphi)}\varphi
\right)\,, 
\]
(The right-hand side of (\ref{antighost}) being linear in the quantum fields
will not get renormalized.)

\noindent -- the Ward identities for the invariances under the
diffeomorphisms (\ref{diff-fields}) and the local Lorentz transformations 
(\ref{lor-fields}): 
\begin{equation}
{\cal W}_{{\rm diff}}\Sigma = \displaystyle{\int} d^3x\,\displaystyle{
\sum_{\Phi}^{}} \delta_{{\rm diff}}^{(\varepsilon)}\Phi {\displaystyle{\frac{
\delta\Sigma }{\delta\Phi }}} = 0 \,,  \label{diffeo}
\end{equation}
and 
\begin{equation}
{\cal W}_{{\rm Lorentz}}\Sigma = \displaystyle{\int} d^3x\,\displaystyle{
\sum_{\Phi}^{}} \delta_{{\rm Lorentz}}^{(\lambda)}\Phi {\displaystyle{\frac{
\delta\Sigma }{\delta\Phi }}} = 0 \,,  \label{localoren}
\end{equation}
where the summations run over all quantum and external fields.

\noindent
The functional operators defined here above, together with the operators 
\begin{equation}
\begin{array}{l}
{\cal G}_a = {\displaystyle{\frac{\delta }{\delta \bar{c}_a}}} +
\partial_\mu \left( eg^{\mu\nu}{\displaystyle{\frac{\delta }{\delta
A_a^{*\nu}}}} \right) \,, \\[3mm] 
{\cal W}_{{\rm rigid}}^a = \displaystyle{\int} d^3x\left( \displaystyle{\
\sum_{\phi =A,c,\bar c,b,A^*,c^*}^{}} f^{\,abc}\phi_b{\displaystyle{\frac{
\delta }{\delta \phi^c}}} +\displaystyle{\sum_{\Phi
=\Psi,\varphi,\Psi^*,\varphi^*}^{}} T^{(\Phi)a} \Phi {\displaystyle{\frac{
\delta }{\delta \Phi}}}\right) \,.
\end{array}
\label{other-op}
\end{equation}
obey the algebra (\ref{fcondnonab1}), (\ref{nilpot1}) and 
\begin{eqnarray}
& {{\displaystyle{\frac{\delta {\cal S}({\cal F})}{\delta b_a}}}} - {\cal B}
_{{\cal F}}\left( {\displaystyle{\frac{\delta {\cal F}}{\delta b_a}}} -
\partial_\mu \left( eg^{\mu\nu}A_\nu^a\right) \right) = {\cal G}^a{\cal \,F}
\,,  \label{fcondnonab3} \\[3mm]
& {\cal G}^a{\cal S}\left( {\cal F}\right) + {\cal B}_{{\cal F}}{\cal G}^a 
{\cal F}=0\,,  \label{fcondnonab4} \\[3mm]
& {\bar{{\cal G}}}^a{\cal S}\left( {\cal F}\right) + {\cal B}_{{\cal F}
}\left( {\bar{{\cal G}}}^a{\cal F} - \Delta_{{\rm cl}}^a\right) ={\cal W}_
{{\rm rigid}}^a {\cal F}\,,  \label{fcondnonab5} \\[3mm]
& {\cal W}_X{\cal S}({\cal F})-{\cal B}_{{\cal F}}{\cal W}_X{\cal F}=0\,,
\qquad X={\rm diff,\,Lorentz\,,\,rigid\,\,\,},  \label{fcondnonab6}
\end{eqnarray}
where ${\cal F}$ is an arbitrary functional of ghost number zero.

One notes that, since the action $\Sigma$ (\ref{total-action}) obeys the
Slavnov-Taylor\ identity (\ref{slavnonab1}) and the gauge condition
(\ref{landau1}), the identities 
(\ref{fcondnonab3}) and (\ref{fcondnonab5}) imply:

\noindent -- the ghost equation of motion 
\begin{equation}
{\cal G}_a\Sigma =0\,,  \label{ghost1}
\end{equation}
\noindent -- the Ward identity expressing the rigid invariance of the
theory, i.e. its invariance under the gauge transformations with constant
parameters, 
\begin{equation}
{\cal W}_{{\rm rigid}}^a\Sigma =0\,.  \label{crigidcondnonab}
\end{equation}
The ghost equation (\ref{ghost1}) implies that the theory depends on the
field $\bar{c}$ and on the antifield $A^{*\mu }$ through the combination 
\begin{equation}
{\hat{A}}_a^{*\mu }=A_a^{*\mu }+eg^{\mu \nu }\partial _\nu \bar{c}_a\,.
\label{combination1}
\end{equation}


\section{Renormalizability}

\label{sect3}

\setcounter{equation}{0} 

We face now the problem of showing that all the constraints defining the
classical theory also hold at the quantum level, i.e., that we can construct
a renormalized vertex functional 
\begin{equation}
\Gamma =\Sigma +{\cal O}\left( \hbar \right) \,\,,  \label{vertex}
\end{equation}
obeying the same constraints and coinciding with the classical action at
order zero in $\hbar$.

As announced in the Introduction, the proof of renormalizability will be
valid for manifolds which are topologically equivalent to a flat manifold
and which admit an asymptotically flat metric. It is only in this case that
one can expand in powers of $\bar{e}_\mu^m$ = $e_\mu ^m$ $-$ $\delta_\mu^m$,
considering $\bar{e}_\mu^m$ as a classical background field in flat 
${\cal R}^3$, and thus make use of the 
general theorems of renormalization theory
actually proved for flat space-time \cite{Zimmerman,qap}. 

\subsubsection*{Power-Counting}

The first point to be checked is power-counting renormalizability. It
follows from the fact that the dimension of the action is bounded by 
three\footnote{The ultraviolet dimension  of the
fields coincide with their canonical dimensions.
They determine the ultraviolet power-counting. If there are massless fields
in the theory, one should take special care of the infrared 
convergence~\cite{low}. 
We shall not insist on this, since we are interested here in the
purely ultraviolet problem of the nonrenormalization of the Chern-Simons
coupling. We shall simply assume that no ``infrared anomaly'' or ``radiative
mass generation''~\cite{mass-gen} occurs.}. 
  The ultraviolet dimension
as well as the ghost number and
the Grassmann parity of all fields and antifields are collected in Table~\ref
{table1}. 
\begin{table}[hbt]
\centering
\begin{tabular}{|c||c|c|c|c|c|c|c|c|c|c|c|}
\hline
& $A_\mu $ & $\Psi $ & $\varphi $ & $b$ & $c$ & ${\overline{c}}$ & $A^{*\mu
} $ & $\Psi ^{*}$ & $\varphi^{*}$ & $c^{*}$ & ${e}_{\mu}^{~a}$ \\ 
\hline\hline
$d$ & 1 & 1 & 1/2 & 1 & 0 & 1 & 2 & 2 & 5/2 & 3 & 0 \\ \hline
$\Phi \Pi $ & 0 & 0 & 0 & 0 & 1 & $-1$ & $-1$ & $-1$ & $-1$ & $-2$ & $0$ \\ 
\hline
$GP$ & 0 & 1 & 0 & 0 & 1 & 1 & 1 & 0 & 1 & 0 & 0 \\ \hline
\end{tabular}
\caption[t1]{ Ultraviolet 
dimension $d$, ghost number $\Phi \Pi $ and Grassmann
parity $GP$.}
\label{table1}
\end{table}

\subsubsection*{Renormalized Functional Identities}

The second point to be discussed is that of the functional identities which
have to be obeyed by the vertex functional.

The gauge condition (\ref{landau1}), ghost equation (\ref{ghost1}),
antighost equation (\ref{antighost}) as well as rigid gauge invariance (\ref
{crigidcondnonab}) can easily be shown to hold at all orders, i.e., are not
anomalous~\cite{pigsor}. The validity to all orders of the Ward identities
of diffeomorphisms and local Lorentz will be assumed in the following: the
absence of  anomalies for them has been proved in 
refs.~\cite{brandt,brandt-bis} 
for the class of manifolds we are considering here.
Therefore we shall be working in the space of diffeomorphism and local
Lorentz invariant functionals.

It remains now to show the possibility of implementing the Slavnov-Taylor\
identity (\ref{slavnonab1}) for the vertex functional $\Gamma$. As it is
well known~\cite{pigsor}, this amounts to study the cohomology of the
nilpotent operator ${{\cal B}_{\Sigma}}$, defined by (\ref{linear}), in the
space of the local integral functionals $\Delta$ of the various fields
involved in the theory. This means that we have to look for solutions of the
form 
\begin{equation}
\Delta = \Delta_{{\rm cohom}} + {{\cal B}_{\Sigma}}\hat\Delta\,,
\label{gen-coho-sol}
\end{equation}
for the equation 
\begin{equation}
{{\cal B}_{\Sigma}}\Delta = 0\,.  \label{cohom-eq}
\end{equation}
$\Delta_{{\rm cohom}}$ represents the cohomology, i.e. the ``nontrivial''
part of the general solution: it cannot be written as a ${{\cal B}_{\Sigma}}
$-variation of a local integral functional $\hat\Delta$.

The solutions $\Delta$ with ghost number 0 represent the arbitrary invariant
counterterms one can add to the action at each order of perturbation theory.
The nontrivial ones correspond to a renormalization of the physical
parameters: coupling constants and masses, whereas the trivial ones
correspond to nonphysical field redefinitions.

The nontrivial solutions with ghost number 1 are the possible gauge
anomalies.

In both cases, power-counting renormalizability restricts the dimension of
the intgrand of $\Delta$ to 3. Moreover, the constraints (\ref{landau1} -- 
\ref{localoren}), (\ref{ghost1}) and (\ref{crigidcondnonab}), valid now for
the vertex functional $\Gamma$, imply for $\Delta$ the conditions 
\begin{equation}
\begin{array}{l}
(1)\quad {\displaystyle{\frac{\delta}{\delta b_a}}}\Delta =0\,,\qquad
(2)\quad \displaystyle{\int} d^3x\,{\displaystyle{\frac{\delta}{\delta \,c^a}
}}\Delta =0\,,\qquad (3)\quad {\cal G}^a\Delta =0\,, \\[3mm] 
(4)\quad {\cal W}_X\Delta =0\,,\qquad X={\rm diff.,\,Lorentz\,,\,rigid}\,.
\end{array}
\label{cond-delta}
\end{equation}

It has been proven in quite generality~\cite{henneaux,brandt-bis} that in
such a gauge theory the cohomology in the sector of ghost number one is
independent of the external fields\footnote{%
The proof is valid for any semi-simple gauge group. But if the gauge group
contains Abelian factors, the cohomology in the sector of ghost number one
may depend on the external fields \cite{henneaux,brandt-bis}. However, it
can be shown~\cite{bbbc} that such terms cannot contribute to the gauge
anomaly due to the Abelian ghost fields being free or at most softly coupled.
}. We can thus restrict the field dependence of $\Delta $ to $A_\mu$, $c$, 
$\varphi$, and $\Psi$ -- the dependence on $c$ being through its
derivatives due to the second of the constraints (\ref{cond-delta}).

Beginning with the anomalies (sector of ghost number 1), one 
knows~\cite{brandt,brandt-bis} that, 
in three dimensions, the cohomology in this sector
is empty, up to possible terms in the Abelian ghosts. They however can be
seen, by using the arguments of~\cite{bbbc}, not to contribute to the
anomaly, due to the freedom or soft coupling of the Abelian ghosts. We thus
conclude to the absence of gauge anomaly, hence to the validity of the
Slavnov-Taylor\ identity (\ref{slavnonab1}) to all orders for the vertex
functional $\Gamma$.

Going now to the sector of ghost number 0, i.e. looking for the arbitrary
invariant counterterms which can be freely added to the action at each
order, we find that the nontrivial part of $\Delta $ admits the general
representation 
\begin{equation}
\Delta _{{\rm phys.}}=\left( z_\kappa \kappa {\displaystyle{\frac \partial
{\partial \kappa }}}+z_\lambda \,\lambda {\displaystyle{\frac \partial
{\partial \lambda }}}+z_y\,y{\displaystyle{\frac \partial {\partial y}}}
\right) \Sigma \,\,,  
\label{cterm11}\end{equation}
where we have only kept the terms of dimension 3 (the lower dimension ones
having being neglected). We have also neglected terms such as $\int
d^3x\,e\,R\Phi ^2$, which do not contribute in the limit of flat space. 
The
coefficients $z_\lambda $ and $z_y$ are invariant tensors, such as the
coupling $\lambda $ and $y$ in (\ref{potential}).

For the trivial part, we find: 
\begin{equation}
{{\cal B}_{\Sigma}} \hat{\Delta} = \left( z_A{\cal N}_A + z_\psi {\cal N}
_\psi + z_\varphi {\cal N}_\varphi \right) \Sigma \,,  
\label{cterm2}\end{equation}
with 
\begin{equation}
\begin{array}{l}
{\cal N}_A \Sigma = 
\left(N_A-N_{A^{*}}-N_b-N_{\bar{c}}\right) \Sigma = 
{\ {\cal B}_{\Sigma}}\displaystyle
{\int} d^3x {\hat A}^{*\mu}_a A_\m^a \,, \\
[3mm] 
{\cal N}_\psi\Sigma =
\left( N_\psi + N_{\bar{\Psi}} - N_{\Psi^{*}} - N_{\bar{
\Psi}^{*}} \right) \Sigma 
= - {{\cal B}_{\Sigma}}\displaystyle{\int} d^3x
\left( {\bar\Psi}^{*A}\Psi_A + 
  {\bar\Psi}^A\Psi^*_A \right)\,, \\[3mm] 
{\cal N}_\vf\Sigma = 
\left( N_\vf - N_{\varphi^{*}} \right) \Sigma = {{\cal B}
_{\Sigma}}\displaystyle{\int} d^3x \varphi^{*i}\varphi_i \,,
\end{array}
\label{count-op}
\end{equation}
where we have introduced the counting operators 
\begin{equation}
N_\Phi =\displaystyle{\int} d^3x\,\Phi {\displaystyle{\frac{\delta}{\delta
\,\Phi }}}\,,\quad \Phi = \mbox{any field}\,.  \label{count-op'}
\end{equation}
Eqs. (\ref{cterm11}) and (\ref{cterm2} -- \ref{count-op}) make manifest the
physical renormalizations, which affect the physical parameters such as the
coupling constants (and the masses), on the one hand, and the nonphysical
renormalizations, which correspond to a mere redefinition of the field
variables, on the other hand.

This concludes the proof of the renormalizability of the model: all
functional identities hold without anomaly and the arbitrary
renormalizations only affect the parameters defined by the initial classical
theory.


\subsubsection*{Callan-Symanzik Equation}

The latter renormalization properties are summarized 
 in the Callan-Symanzik
equation, which results from the expansion of the invariant insertion
defined by 
\[
{\cal D}\Gamma=\sum\limits_{{\rm all\ dimensionful\ parameters\ \mu}} \mu 
\frac{\partial\Gamma} {\partial \,\mu }\,\,, 
\]
in the quantum basis defined by (\ref{cterm11}) and (\ref{cterm2}) -- 
$\Sigma $ being replaced there by $\Gamma$: 
\begin{equation}
\left( {\cal D} + \beta _\kappa \partial _\kappa +\beta_y\partial_y+\beta
_\lambda \partial _\lambda - \gamma_A{\cal N}_A-\gamma_\psi {\cal N}_\psi-
\gamma_\vf{\cal N}_\varphi \right) \Gamma\, \sim 0 \,,
\label{callan-symanzik}
\end{equation}
where $\sim $ means equality up to mass terms and dimensionful couplings,
the terms contribuiting which vanish in the flat limit being neglected, too.

The aim of the next and last section is to prove that $\beta_\kappa$ is
vanishing.


\section{Nonrenormalization of the Chern-Simons \protect\\
Coupling}\label{sect4}

\setcounter{equation}{0} 

In order to get more deeply into the scaling properties of the present
theory, we need a local form of the Callan-Symanzik equation. This will
allow us to exploit the fact that the integrand of the Chern-Simons term in
the action is not gauge invariant, although its integral is. Such a local
form of the Callan-Symanzik equation is provided by the 
``trace identity''.

In order to derive the latter, let us first introduce the 
energy-momentum
tensor, defined as the following tensorial quantum insertion obtained 
as the
derivative of the vertex functional with respect to the dreibein: 
\begin{equation}
{\Theta }_\nu ^{~\mu }\cdot \Gamma = e^{-1} e_\nu ^{~m}~\frac {\delta
\Gamma}{\delta e_\mu ^{~m}}\,\,.  \label{theta}
\end{equation}

   From the diffeomorphism Ward identity (\ref{diffeo}), follows the 
covariant
conservation law of the energy-momentum tensor: 
\begin{equation}
e\,\nabla _\mu \left[ {\Theta }_\nu ^{~\mu }(x)\cdot \Gamma \right] 
= {w}_\nu \left( x\right) \Gamma +
\nabla _\mu {w}_\nu ^{~\mu }\left(x\right) \Gamma \,,  
\label{cons-theta}
\end{equation}
where $\nabla _\mu $ is the covariant derivative with respect to the
diffeomorphisms. The right hand side is an equation of motion, the
functional differential operators $w_\lambda \left( x\right) $ and 
$w_\nu
{}^\mu \left( x\right) $ being given by (\ref{a9}) and (\ref{a10}),
respectively.

The trace $\Theta _\mu ^{~\mu}(x)\cdot\Gamma$ 
turns out to be vanishing, up
to total derivatives, mass terms and dimensionful couplings, in the
classical approximation, due to the field equations, which means that (\ref
{theta}) is the improved energy-momentum tensor. This is shown in 
Appendix
A where we have indeed derived, for the classical theory, the equation 
\begin{equation}
w\left( x\right) \Sigma \sim \partial _\mu \,\Lambda ^\mu \left( x\right)
\,\,,
\label{class-tr1}\end{equation}
or, equivalently:
\begin{equation}
e \Theta_\mu{}^\mu(x) \sim  
  w^{\rm trace}(x)\Sigma + \partial_\mu\Lambda^\mu(x)\ ,
\label{class-tr2}\end{equation}
with
\begin{equation}
w = e_\mu^{~m} \frac{\delta}{\delta e_\mu^{~m}} - w^{{\rm trace}}\ ,
\label{a15'}\end{equation}
\begin{equation}
w^{{\rm trace}} =\frac 12\varphi \frac \delta {\delta \varphi
}-\frac 12\varphi ^{*}\frac \delta {\delta \varphi ^{*}}+\Psi \frac \delta
{\delta \Psi }+\bar{\Psi}\frac \delta {\delta \bar{\Psi}}-\Psi ^{*}\frac
\delta {\delta \Psi ^{*}}-\bar{\Psi}^{*}\frac \delta {\delta \bar{\Psi}^{*}}
+{\ \bar{c}}^a\frac \delta {\delta {\bar{c}}^a}+b^a\frac \delta {\delta
b^a}\,\,,  
\label{a16}\end{equation}
and 
\begin{equation}
\Lambda ^\mu = e\,i\bar{\Psi}\gamma ^\mu \Psi 
  +e \varphi\nabla^\mu \varphi
  -s\,\left( e\,g^{\mu \nu }\bar{c}A_\nu \right) \,\,.
\label{a17}\end{equation}  
The symbol $\sim $ in (\ref{class-tr1}) and (\ref{class-tr2})
again means equality up to mass terms and
dimensionful couplings.

Let us now look for the quantum version of the trace identity 
(\ref{class-tr1}) or (\ref{class-tr2}).
We first observe that the following commutation relations hold:
\begin{eqnarray}
\left[ \frac \delta {\delta b^a\left( y\right) },w\left( x\right) \right]
=-\delta \left( x-y\right) \frac \delta {\delta b^a\left( x\right)
}\,\,,\,\,\,\,\,\,\,\,\,\,\,\,\,\,\,
\left[ {\bar{{\cal G}}}^a , w\left( x\right) \right] =0
\,\,,\,  \nonumber \\[3mm] 
\left[{\cal G}^a\left( y\right) ,w\left( x\right) \right]=-\delta
\left( x-y\right) {\cal G}^a\left( x\right) 
+\partial_\mu \delta \left(x-y\right) 
\left(eg^{\mu \nu }\frac \delta {\delta A_a^{*\nu }}\right)(y)
\,\,.  
\label{abc}\end{eqnarray}
Let us then define the quantum extension $\Lambda^\mu\cdot\Gamma$ 
of the classical expression (\ref{a17}) as a vector insertion obeying
-- beyond the conditions of invariance or covariance
under ${\cal B}_\Gamma$, 
${\cal W}_{\rm diff}$, ${\cal W}_{\rm Lorentz}$ and 
${\cal W}_{\rm rigid}$ -- the constraints
\begin{eqnarray}
\frac \delta {\delta b^a\left(y\right) }\left[ \,\Lambda ^\mu \left(
x\right) \cdot \Gamma \right] &=&-\delta \left( x-y\right) e\,g^{\mu \nu
}A_\nu^a  \,\,,\,\,\,\,\,\,\,\,\,\,\,\,\,\,\,
{\bar{{\cal G}}}^a \left[ \,\Lambda ^\mu \left( x\right) \cdot \Gamma 
\right] =0\,\,,  \nonumber \\ [3mm]
{\cal G}^a\left( y\right) \left[\,\Lambda ^\mu \left( x\right) 
\cdot \Gamma \right] &=&\delta \left( x-y\right) eg^{\mu \nu }
 \frac{\delta\Gamma }{\delta A_a^{*\nu }}\,\,,  
\label{abc'}\end{eqnarray}
already obeyed in the classical limit. Then the divergence
$\partial_\m\left[\Lambda^\mu\cdot\Gamma\right]$ will obey the same  
constraints as $\omega(x)\Gamma$, which read 
\begin{eqnarray}
\frac \delta {\delta b^a(y) } w(x)\Gamma
  &=&-\partial_\mu\delta(x-y) \left(e\,g^{\mu\nu}A^a_\nu\right)(y)
\,\,,\,\,\,\,\,\,\,\,\,\,\,\,\,\,\,
{\bar{{\cal G}}}^a w(x)\Gamma =0 \,\,,\,  \nonumber \\[3mm] 
{\cal G}^a(y) w(x)\Gamma &=& \partial_\mu\delta(x-y)
 \left(e\,g^{\mu \nu}\frac{\delta\Gamma}{\delta A_a^{*\nu }}\right)(y)
\,\,,
\label{constr-omega}\end{eqnarray}
and follow from the commutation relations (\ref{abc}).

Thus, the dimension 3 insertion $\Delta\cdot\Gamma$ defined by
\begin{equation}
w(x) \Gamma \sim
\partial_\mu \left[ \,\Lambda^\mu (x) \cdot \Gamma \right] 
+  \Delta(x)\cdot\Gamma          \,\,,  
\label{quant-trace}\end{equation}
-- which expresses the quantum corrections to the equation 
(\ref{class-tr1}) --
beyond of being invariant or covariant
under ${\cal B}_\Gamma$, 
${\cal W}_{\rm diff}$, ${\cal W}_{\rm Lorentz}$ and 
${\cal W}_{\rm rigid}$,  
will obey the homogeneous constraints
\begin{equation}
{{\frac \delta {\delta b^a(y)}}}\left[\Delta(x)\cdot\Gamma\right] = 0\,,\qquad 
{\bar{{\cal G}}}^a\left[\Delta(x)\cdot\Gamma\right] = 0\,,\qquad 
{\cal G}^a(y)\left[\Delta(x)\cdot\Gamma\right] = 0\,\,\,.  
\label{cond1}\end{equation}
We have to expand $\Delta\cdot\Gamma$ in a basis of invariant insertions 
 satisfying the constraints 
(\ref{cond1}), 
keeping only terms of dimension 3. Such a basis is given in the
classical approximation by 
\begin{equation}
\begin{array}{l}
\left\{ \varphi ^6~,~~\varphi ^2\bar{\Psi}\Psi ~,
~~e^{-1}{\cal B}_\Sigma ({
\hat{A}}^{*}A)~,~~e^{-1}{\cal B}_\Sigma 
  (\bar{\Psi}^{*}\Psi+\bar{\Psi}\Psi^{*})~,
~~e^{-1}{\cal B}
_\Sigma (\varphi ^{*}\varphi )~\right\} \,,
\end{array}
\label{class-basis}\end{equation}
up to total derivative terms which may be reabsorbed in the insertion
$\Lambda^\mu$.

\noindent {\it Note}: We have separated out 
explicitly the cohomologically trivial
elements (i.e. ${\cal B}_\Sigma $-variation) from the nontrivial ones, and
only the terms surviving in the flat limit are kept.

An appropriate quantum extension of this basis is 
\begin{equation}
\begin{array}{l}
\left\{ e^{-1}n_\lambda \cdot \Gamma ~,~~e^{-1}n_y\cdot \Gamma
~,~~e^{-1}n_A\cdot \Gamma ~,~~e^{-1}n_\psi \cdot \Gamma ~,~~e^{-1}n_\varphi
\cdot \Gamma ~\right\} \,,
\end{array}
\label{quantum-basis}\end{equation}
up to total derivative terms which may be reabsorbed in 
a redefinition of the insertion
$\Lambda^\mu\cdot\Gamma$.
Here, $n_\lambda $ and $n_y$ are the integrands of the insertions obtained
by differentiating the vertex functional with respect to the couplings 
$\lambda $ and $y$, respectively: 
\[
\begin{array}{l}
n_\lambda \cdot \Gamma ~:\quad \displaystyle{\int }d^3x~n_\lambda \cdot
\Gamma ={\ \displaystyle{\frac{\partial \Gamma }{\partial \lambda }}}\,, \\
[3mm] 
n_y\cdot \Gamma ~:\quad \displaystyle{\int }d^3x~n_y\cdot \Gamma =
{\ \displaystyle{\frac{\partial \Gamma }{\partial y}}}\,.
\end{array}
\]
These integrands -- defined up to total derivatives -- are choosen such as
to satisfy the constraints (\ref{cond1}).

\noindent An obvious choice for $n_\psi $ and $n_{\varphi}$ is: 
\[
\begin{array}{l}
n_\psi \cdot \Gamma =-{\cal B}_\Gamma (\bar{\Psi}{}^{*}\Psi +\bar{\Psi}\Psi
^{*})=\bar{\Psi}{\displaystyle{\frac{\delta \Gamma }{\delta \bar{\Psi}}}}- 
\bar{\Psi}{}^{*}{\displaystyle{\frac{\delta \Gamma }{\delta \bar{\Psi}{}^{*}}
}}+\mbox{ conjugate}\,, \\[3mm] 
n_{\varphi}\cdot \Gamma ={\cal B}_\Gamma (\varphi ^{*}\varphi )=\varphi 
{\ \displaystyle{\frac{\delta \Gamma }{\delta \varphi }}}-\varphi ^{*}
{\ \displaystyle{\frac{\delta \Gamma }{\delta \varphi ^{*}}}}\,.
\end{array}
\]
Their integrals indeed yield the counting operators ${\cal N}_\psi$ and 
${\cal N}_\varphi$ defined by (\ref{count-op}) where $\Sigma$ is replaced
by $\Gamma $. Less obvious is the definition of $n_A$, due to the fact that 
$\hat{A}{}^{*}A$ is not linear in the quantum fields 
(see (\ref{combination1})). In order to define it properly, 
keeping both properties 
satisfied by its
classical counterpart, i.e. its BRS invariance and the fact that its
integral reproduce the counting operator ${\cal N}_A$, we use a point
splitting regularization ($x^\mu $ $\longrightarrow $ $x^\mu \pm \epsilon
^\mu$): 
\begin{equation}
n_A\cdot \Gamma ={\cal B}_\Gamma (m_A\cdot \Gamma )={\cal B}_\Gamma \left\{
\lim_{\epsilon \to 0}{\hat{A}}^{*}(x+\epsilon )A(x-\epsilon )\right\} \,.
\label{splitting}
\end{equation}
That this limit exists is shown in Appendix B.

Thus, the expansion of $\Delta\cdot\Gamma$ 
in the basis (\ref{quantum-basis}) we have just constructed
yields, after substitution into (\ref{quant-trace}) and use of
(\ref{a15'}), the local trace identity in curved space-time: 
\beq\ba{l}
e\,\Theta _\mu ^{~\mu }\left( x\right) \cdot \Gamma \sim \es
\qquad \left\{ \beta_{\lambda \cdot }n_\lambda +
\beta _yn_y-\gamma _An_A-\gamma _\psi n_\psi
-\gamma _\varphi n_\varphi \right\} \cdot \Gamma 
+\,w^{{\rm trace}}\left( x\right) \cdot \Gamma +\partial _\mu 
\,\left[ \Lambda ^\mu \left( x\right) \cdot \Gamma \right] \es
\qquad + \mbox{\rm terms vanishing in the flat limit}\,.  
\ea\eqn{trace-id-tilde1}

In order to make the connection between the trace of the energy-momentum
tensor and the Callan-Symanzik equation, lets us consider a while the limit
of flat space-time, where rigid dilatation symmetry makes sense. In this
limit, (\ref{cons-theta}) and 
(\ref{trace-id-tilde1}) hold with $e=1$ and $\nabla _\mu 
$ $=$ $\partial _\mu $. We can therefore define the dilatation current as 
\begin{equation}
\left[ {\cal D}^\mu \cdot \Gamma \right] _{{\rm flat}}=x^\nu \left[ {\Theta}
_\nu ^{~\mu }\left( x\right) \cdot \Gamma \right] _{{\rm flat}}-x^\nu \left[
w_\nu ^{~\mu}\left( x\right) \cdot \Gamma \right] _{{\rm flat}}-
\left[\Lambda^{\mu }\left( x\right) \cdot \Gamma \right] _{{\rm flat}}\,\,,
\label{dilacurr}
\end{equation}
which, in the classical approximation, is conserved up to mass terms and
dimensionful couplings (see (\ref{dilata1})). Thus, for the renormalized
theory, in the flat limit, we can write as usually the Callan-Symanzik
equation -- which is the Ward identity for anomalous dilatation invariance
-- as the integral trace identity 
\begin{equation}
\displaystyle{\int }d^3x~\left[ \Theta _\mu ^{~\mu }\left( x\right) \cdot
\Gamma \right] _{{\rm flat}}\sim \left( \beta _\kappa \partial _\kappa
+\beta _y\partial _y+\beta _\lambda \partial _\lambda -\gamma _A{\cal N}
_A-\gamma _\psi {\cal N}_\psi -\gamma _\varphi {\cal N}_\varphi \right)
\cdot \Gamma _{{\rm flat}}\,.  \label{flat-trace}
\end{equation}
This observation allows us to identify
the coefficients $\beta $ and $\gamma $ of the expansion 
(\ref{trace-id-tilde1}) with those of the Callan-Symanzik equation 
(\ref{callan-symanzik}). 
In particular, the absence of a term corresponding to
the integrand of the Chern-Simons action in the 
basis (\ref{quantum-basis}) of invariant local
operators proves the vanishing of the Chern-Simons $\beta $ function: 
\begin{equation}
\beta _\kappa =0\,.  
\label{beta=0}\end{equation}


\section{The Yang-Mills Chern-Simons Theory in Curved Space-Time}

\label{sect5}

\setcounter{equation}{0} 

The  YMCS action, invariant under the diffeomorphisms and local
Lorentz transformations, and gauge invariant -- i.e. BRS-invariant 
 -- in the Landau gauge, is of the form 
\begin{eqnarray}
\Sigma_{{\rm inv}}+\Sigma_{{\rm gf}} &=&\int d^3x\,\left\{ -\frac e4F_{\mu
\nu }^aF^{a\mu \nu }+m\,\varepsilon ^{\mu \nu \rho }\left( A_\mu ^a\partial
_\nu A_\rho ^a+\frac g3f_{abc}A_\mu ^aA_\nu ^bA_\rho ^c\right) \right\} 
\nonumber \\
&&-\,{\int }d^3x\,\,e\,g^{\mu \nu }\left( \partial _\mu b_aA_\nu ^a+\partial
_\mu \bar{c}_aD_\nu c^a\right) \,\,,  \label{g-inv-action}
\end{eqnarray}
where $m$ is a dimensionful coupling constant --  the topological
mass indeed~\cite{deser} -- 
and $g$ is the gauge coupling constant,  dimensionful, too.

The field strength and covariant derivative are defined 
as\footnote{{}The parametrization of 
the coupling constant, 
differing from the one used in the preceding sections, 
is adapted to the fact that it is now dimensionful.}  
\begin{equation}
F^a_{\mu \nu }=\partial _\mu A^a_\nu -\partial _\nu A^a_\mu 
 + g f _{abc} A^b_\mu A^c_\nu  \,\,,
\end{equation}
\begin{equation}
D_\mu c^a=\partial _\mu c^a +   g f _{abc} A^b_\mu c^c \,\,.
\end{equation}

The BRS transformations and antighost equation now read
\begin{equation}
\begin{tabular}{cc}
$sA_\mu ^a=-D_\mu c^a\,,$ & $sc^a={\displaystyle}\frac g2f_{bc}{}^ac^bc^c\,,$
\\ [3mm] 
$s{\bar{c}}^a=b^a\,,$ & $sb^a=0\,.$
\end{tabular}
\label{BRSab}
\end{equation}
and  
\begin{equation}
{\bar{{\cal G}}}^a\Sigma \,\,=\,\, 
\displaystyle{\int} d^3x
 \left({\displaystyle{\ \frac{\delta}{\delta c^a}}}
 + gf^{\,abc}\bar{c}_b {\displaystyle{\frac{\delta}{\delta b^c}}}\right)
   \Sigma\,\,=\,\, 
\Delta_{{\rm cl}}^a\,,  
\label{ym-antighost}\end{equation}
with 
\[
{\bar{{\cal G}}}^a\,\,=\,\,\displaystyle{\int }d^3x\left( {\displaystyle{\
\frac \delta {\delta c^a}}}+g\,f^{\,abc}\bar{c}_b{\displaystyle{\frac \delta
{\delta b^c}}}\right) \,\,,\qquad \Delta _{{\rm cl}}^a=g\displaystyle{\int}
d^3x\,f^{\,abc}\left( A_b^{*\mu }A_{c\mu }-c_b^{*}c_c\right) \,,
\]

The functional identities (\ref{linear}), (\ref{diffeo}), (\ref{localoren})
and the constraints (\ref{landau1}), (\ref{ghost1}) and (\ref
{crigidcondnonab}) are unchanged -- except  for
the matter terms which are now absent.

Finally, in order to quantize the system (\ref{g-inv-action}) we add an
action term $\Sigma _{{\rm ext}}$ for the coupling of the BRS
transformations to external fields: 
\begin{equation}
\Sigma _{{\rm ext}}=\displaystyle{\int }d^3x\sum_{\Phi =A_\mu ^a,\,c^a}\Phi
^{*}s\Phi \,.  \label{ext-action1}
\end{equation}


\section{Renormalizability and Quantum Scale Invariance}

\label{sect6}

\setcounter{equation}{0} 


\subsubsection*{Power-Counting}


The first point to be checked is power-counting renormalizability -- 
 in fact
superrenormalizability. It follows from  the dimension of the
action being bounded by three. 
The ultraviolet dimension, as well as the ghost
number and the Grassmann parity of all fields and antifields are collected
in Table~\ref{table2}.

\begin{table}[tbh]
\centering
\begin{tabular}{|c||c|c|c|c|c|c|c|}
\hline
& $A_\mu $ & $b$ & $c$ & ${\overline{c}}$ & $A^{*\mu }$ & $c^{*}$ & $g$ \\ 
\hline\hline
$d$ & $1/2$ & $3/2$ & $-1/2$ & $3/2$ & $5/2$ & $7/2$ & $1/2$ \\ 
\hline 
$\Phi\Pi $ & $0$ & $0$ & $1$ & $-1$ & $-1$ & $-2$ & $0$ \\ 
 \hline 
$GP$ & 0 & 0 & 1 & 1 & 1 & 0 & 0 \\ \hline 
\end{tabular}
\caption[t1]{ Ultraviolet
dimension $d$, ghost number $\Phi \Pi $ and Grassmann
parity $GP$.}
\label{table2}
\end{table}

In order to explicitely find the possible renormalizations and anomalies of
the theory, we can use the following result~\cite{mpr}: The degree of
divergence of a 1-particle irreducible Feynman graph $\gamma$ is given by 
\begin{equation}
d\left( \gamma \right) =3-\sum\limits_\Phi d_\Phi N_\Phi -\frac 12N_g\,\,.
\label{power}
\end{equation}
Here $N_\Phi$ is the number of external lines of $\gamma$ corresponding to
the field $\Phi $, $d_\Phi $ is the dimension of $\Phi$ as given 
in Table 2, and $N_g$ is the power of the coupling constant $g$ in 
the integral
corresponding to the diagram $\gamma $. The dependence on the coupling
constant is characteristic of a superrenormalizable theory.

The equivalent expression 
\begin{equation}
d\left( \gamma \right) =4-\sum\limits_\Phi \left( d_\Phi +\frac 12\right)
N_\Phi -L\,\,,
\end{equation}
where $L$ is the number of loops of the diagram, shows that only graphs up
to two-loop order are divergent.

In order to apply the known results on the quantum action 
principle~\cite{qap} to the present situation, 
one may consider $g$ as an external field of
dimension $\frac 12$. Including it in the summation under $\Phi$, (\ref
{power}) gets the same form as in a strictly renormalizable theory: 
\begin{equation}
d\left( \gamma \right) =3-\sum\limits_{\tilde{\Phi}=\Phi ,\,g}d_{\tilde{\Phi}
}N_{\tilde{\Phi}}\,\,,\,\,\,\,\,\mbox{with}\,\,\,\,\,d_g=\frac 12\,\,.
\end{equation}

Thus, including the dimension of $g$ into the calculation, we may state that
the dimension of the counterterms of the action is bounded by 3. But, since
they are generated by loop graphs, they are of order 2 in $g$ at least. This
means that, not taking now into account the dimension of $g$, we can
conclude that their real dimension is bounded by 2. The same holds for the
possible breakings of the Slavnov-Taylor identity.

The absence of anomaly for the Slavnov-Taylor identity holds here in the
same way as previously discussed, in Sect.3.

 Let us now look for the arbitrary
invariant counterterms which can be freely added to the action at each
order. According to  the above discussion the counterterm 
is at least of order 
$g^2$. Thus, the most general expression 
for  the nontrivial part of $\Delta$ reads 
\begin{equation}
\Delta _{{\rm phys.}}=z_m\int d^3x\,m\,\varepsilon ^{\mu \nu \rho }\left(
A_\mu ^a\partial _\nu A_\rho ^a+\frac g3f_{abc}A_\mu ^aA_\nu ^bA_\rho
^c\right) \,\,,  \label{cterm0}
\end{equation}
where $z_m$ is arbitrary parameter. Expression (\ref{cterm0}) admits the
suitable representation 
\begin{equation}
\Delta _{{\rm phys.}}=z_mm\frac \partial {\partial m}\Sigma \,\,,
\label{cterm1}
\end{equation}
where $\Sigma$ is the classical action (\ref{total-action}). Eq. (\ref
{cterm0}) shows that a priori only the parameter $m$ can get radiative
corrections. This means that the $\beta $-function related to the gauge
coupling constant $g$ is vanishing to all orders 
of  perturbation theory,
and the anomalous dimensions of the fields as well. We can just state that
the radiative corrections can be reabsorbed through a redefinition of the
topological mass only. This concludes the proof of the renormalizability of
the theory: all functional identities hold without anomaly and the 
renormalizations  might only affect the  the Chern-Simons coupling, i.e. 
the topological mass  $m$.  But the latter turns out to be not
renormalized, too.  We shall indeed show in the next Section that the 
corresponding $\beta$-function vanishes.


\subsubsection*{Quantum Scale Invariance}


The argument is very similar to the one presented in Sect. \ref{sect4}.
However, the equations (\ref{class-tr1} -- \ref{a17}) 
for the classical theory are now replaced by\footnote{See appendix A.}
\beq
w(x)\Sigma \equiv \left( e_\mu^a(x) \frac{\delta}{\delta e_\mu^a(x)}
 - w^{{\rm trace}}(x) \right) \Sigma = \Lambda(x)\ ,
\eqn{ym-cl-tr1}
or, equivalently:
\begin{equation}
e \Theta_\mu{}^\mu(x) =
  w^{\rm trace}(x)\Sigma + \Lambda(x)\ ,
\label{ym-cl-tr2}\end{equation}
with 
\begin{equation}
w^{{\rm trace}}\left( x\right) =
-{1\over2}\left( A^a_\mu \frac \delta {\delta A^a_\mu}-
A_a^{*\mu }\frac \delta {\delta A_a^{*\mu }}+c^a\frac \delta {\delta
c^a}-c_a^{*}\frac \delta {\delta c_a^{*}}\right) +\frac 32{\ }\left({\bar{c}}
^a\frac \delta {\delta {\bar{c}}^a}+b^a\frac \delta {\delta b^a}\right) \,\,,
\label{a20}\end{equation}
and
\beq
\Lambda = 
\dfrac{m}{2}\varepsilon ^{\mu \nu \lambda }A_\mu ^aF_{\nu \lambda }^a
- \dfrac{g}{2} f_{abc}\left( e\,F^{\mu \nu \,a}A_\mu ^bA_\nu ^c
+ {\hat{A}}^{*\mu \,a}A_\mu^bc^c -\dfrac{1}{2} c^{*\,a}c^ bc^c \right)
 + \mbox{ total derivative terms}\ ,
\eqn{lambda}
$\Lambda$ being invariant under ${\cal B}_\Sigma$. 

\noindent {\it Note}: The latter is the effect of the breaking of 
scale invariance due to the dimensionful couplings. The dimension of 
$\Lambda$ -- the dimensions of $g$ and $m$ not being taken
into account -- is lower than three: it is a soft breaking.

The commutation
relations \equ{abc} are changed accordingly into:
\begin{eqnarray}
\left[ \frac \delta {\delta b^a\left( y\right) },w\left( x\right) \right] 
&=&-\frac 32\delta \left( x-y\right) \frac \delta {\delta b^a\left( 
x\right) }\,\,,  \nonumber \\  [3mm]
\left[ {\cal G}^a\left( y\right) ,w\left( x\right) \right] 
&=&-\frac 32\delta \left( x-y\right) {\cal G}^a\left( x\right) 
  +\frac 32\partial_\mu \delta(x-y) 
    \left( eg^{\mu\nu} \frac \delta {\delta A_a^{*\nu}} \right)(y) \,\,,  
\label{cba} \\ [3mm]
\left[ {\bar{{\cal G}}}^a ,w(x) \right]  
&=& \frac 12  \frac \delta {\delta c^a(x) }\,\,.\,  \nonumber
\end{eqnarray}

Now the relations (\ref{cba}) applied to the vertex functional $\Gamma$
yield for insertion $w(x)\Gamma$ the properties 
\begin{eqnarray}
\frac \delta {\delta b_a(y) } w(x) \Gamma  
&=&-\frac 32 \pa_\m\delta(x-y) \left(e\,g^{\mu\nu}A^a_\nu\right)(y) \,\,,
\nonumber \\ [3mm]
{\cal G}^a(y) w(x)\Gamma 
&=&\frac 32\pa_\m\delta(x-y)
  \left( eg^{\mu\nu}\frac{\delta \Gamma }{\delta A_a^{*\nu}}\right)(y) \,\,, 
\label{cba'} \\[3mm]
{\bar{{\cal G}}}^a w(x)\Gamma &=& \frac 12 
\frac {\delta\Gamma} {\delta c_a(x)}
\,\,.  \nonumber
\end{eqnarray}
where we again use the fact that the constraints (\ref{landau1}), (\ref
{antighost}) and (\ref{ghost1}) can be maintained at the quantum level.

The quantum version of \equ{ym-cl-tr1} or \equ{ym-cl-tr2} will be
written as
\begin{equation}
w(x) \Gamma = 
\Lambda(x) \cdot \Gamma +  \Delta(x)\cdot\Gamma          \,\,,  
\label{ym-quant-trace}\end{equation}
where $\Lambda(x)\cdot\Gamma$ is some quantum extension
of the classical insertion \equ{lambda}, 
subjected to the same constraints \equ{cba'}
as $w(x)\Gamma$. It follows that the insertion $\Delta\cdot\Gamma$
defined by \equ{ym-quant-trace} obeys the homogeneous constraints 
\equ{cond1}, beyond of the usual invariances or covariances.
The power-counting dimension of $\Delta$ is 3, but being an effect of
the radiative corrections, it possess a factor $g^2$ at least, and thus
its effective dimension is at most two -- let us recall that we have
attributed power-counting dimension $1/2$ to $g$. It turns out that 
there is no 
insertion obeying all these constraints  -- the Yang-Mills Lagrangian 
density has terms of dimension 3 (without factor $g^2$)
and the Chern-Simons Lagrangian 
density is not BRS invariant.
Hence $\Delta\cdot\Gamma=0$: 
there is no radiative correction to the insertion 
$\Lambda\cdot\Gamma$ describing the breaking of scale invariance, 
and \equ{ym-quant-trace} becomes
\begin{equation}
e\,{\Theta }_\mu ^{~\mu }(x)\cdot\Gamma 
= w^{{\rm trace}}( x) + \Lambda(x)\cdot \Gamma \,.  
\label{trace-id-tilde}
\end{equation}

As in Sect. \ref{sect4}, this local trace identity leads to a
Callan-Symanzik equation:
\beq {\cal D}\Gamma = \int d^3x \Lambda(x)\cdot\Gamma\,,
\end{equation}
but now with no radiative effect at all: the
$\beta$-functions associated to the parameters $g$ and $m$ both 
vanish,
scale invariance remaining affected only by the soft breaking
$\Lambda$.


\section{Conclusions}

\setcounter{equation}{0} 
We have thus achieved the proof of absence of renormalization of the
Chern-Simons coupling (Eq. (\ref{beta=0})) for the 
Chern-Simons theory coupled with
matter through a direct use of the nongauge invariance of the Chern-Simons
three-form. As a by-product, we have obtained a trace identity valid in
curved space-time (Eq. (\ref{trace-id-tilde1})), equivalent to a local
version of the Callan-Symanzik equation (\ref{callan-symanzik}), the result 
(\ref{beta=0}) being taken into account. The same techniques has allowed us
to give a simple proof of the finiteness of the 
Yang-Mills -- Chern-Simons theory. 


\section*{Acknowledgements}

We thank Sylvain Wolf for a useful critical reading of the manuscript,
and the authors of Refs.~\cite{silvio,silvio2} for interesting
discussions on the subject of the present paper and on their work.
This work has been done in parts at the {\it CBPF-DCP}, at the 
{\it Dept. of Physics} of the
{\it UFMG} and at the the {\it Dept. of Physics} of the {\it UFES}. 
We thank all these three
institutions for their financial help for travel expenses and for their kind
hospitality. One of the authors (O.M.D.C.) 
dedicates this work to his wife, Zilda Cristina, to his daughter, Vittoria, 
and to his son, Enzo, who is coming.

\appendix 
\renewcommand{\theequation}{\Alph{section}.\arabic{equation}}
\renewcommand{\thesection}{\Alph{section}}

\setcounter{equation}{0} \setcounter{section}{1}


\section*{Appendix A: Trace Identity and Dilatation Invariance \protect\\
for the Classical Theory}


\subsection*{Chern-Simons Theory coupled to Matter:}

The conservation of the energy-momentum tensor $\Theta _\lambda ^{~\mu }$ is
a consequence of the diffeomorphism Ward Identity (\ref{diffeo}) and of the
definition (\ref{theta}), yielding the following local equation 
\begin{equation}
e\,\nabla _\mu \Theta _\lambda ^{~\mu }\left( x\right) =w_\lambda \left(
x\right) \Sigma +\nabla _\mu w_\lambda ^{~\mu }\left( x\right) \Sigma \,\,,
\label{a1}
\end{equation}
where $\nabla _\mu $ is the covariant derivative with respect to the
diffeomorphisms\footnote{For a tensor $T$, such as, e.g., $A_\m$ or 
${\delta}/{\delta A^{*\mu}} $: 
\[
\nabla_\l T^{\mu\cdots}_{\nu\cdots} = \partial_\l T^{\mu\cdots}_{\nu\cdots}
+ \Gamma_\l{}^\m{}_\rho T^{\rho\cdots}_{\nu\cdots} + \cdots - \Gamma_\l{}^ 
\rho{}_\n T^{\mu\cdots}_{\rho\cdots} - \cdots \,, 
\]
where the $\Gamma_.{}^.{}_.$'s are the Christoffel symbols corresponding to
the connexion $\omega_\m^{mn}$. The covariant derivative of a tensorial
density ${\cal T}$, such as, e.g., $A^{*\mu}$ or 
$\delta/{\delta A_\m}$, is
related to that of the tensor $e^{-1}{\cal T}\,$ by: 
\[
\nabla_\l {\cal T}^{\mu\cdots}_{\nu\cdots} = e\,\nabla_\l\left( e^{-1}
{\cal T}^{\mu\cdots}_{\nu\cdots} \right)\,. 
\]
}, with the differential operators $w_\lambda \left( x\right)$ and 
$w_\lambda ^{~\mu }\left( x\right) $ acting on $\Sigma$ representing contact
terms: 
\begin{equation}
w_\lambda \left( x\right) =\sum\limits_{{\rm all\,fields}}\left( \nabla
_\lambda \Phi \right) \frac \delta {\delta \Phi }\,\,,  \label{a9}
\end{equation}
(becoming the translation Ward operator in the limit of flat space), and 
\beq\ba{l}
w_\lambda ^{~\mu }\left( x\right)=\es
\qquad A^{*\mu }\dfrac{\delta}{\delta A^{*\lambda
}}-A_\lambda \dfrac{\delta}{\delta A_\mu }-\delta _\lambda ^{~\mu }\,\left[
\varphi ^{*}\dfrac{\delta}{\delta \varphi ^{*}}+
 \Psi ^{*}\dfrac{\delta}{\delta
\Psi ^{*}}+\bar{\Psi}^{*}\dfrac{\delta}{\delta \bar{\Psi} ^{*}}+c^{*}
\dfrac{\delta}{\delta c^{*}}+A^{*\mu }
\dfrac{\delta}{\delta A^{*\mu }}\right]
\,\,.  
\ea\eqn{a10}
The integral of the trace of the tensor $\Theta _\lambda ^{~\mu }$, 
\begin{equation}
\displaystyle{\int }d^3x\,e\,\Theta _\mu^{~\mu }=\displaystyle{\int}
d^3x\,e_\mu ^{~m}\frac{\delta \Sigma }{\delta e_\mu ^{~m}}
\equiv N_e\Sigma \,\,,
\label{int-theta}
\end{equation}
turns out to be an equation of motion -- which means that $\Theta _\lambda
^{~\mu }$ is the improved energy-momentum tensor. This follows from the
identity, which is easily checked by inspection of the classical action 
\begin{equation}
N_e\Sigma \sim \left( N_\psi +N_{\bar{\Psi}}+\frac 12N_{\varphi}+N_b+N_{\bar{
c}}-N_{\Psi ^{*}}-N_{{\bar{\Psi}}^{*}}-\frac 12N_{\varphi ^{*}}\right)
\Sigma \,,  \label{int-theta1}
\end{equation}
where $\sim$ means equality up to mass terms and dimensionful
couplings, and the $N_\Phi $'s are the counting operators defined by (\ref
{count-op'}). $N_e$ is the counting operator of the dreibein 
field $e_\mu^m$. 
It is interesting to note that (\ref{int-theta}) is nothing but the Ward
identity for rigid Weyl symmetry~\cite{iorio} -- broken by the mass terms and
dimensionful couplings.

Considering the integrand of (\ref{int-theta1}), taking (\ref{int-theta})
into account, we may write: 
\begin{equation}
w\left( x\right) \Sigma \sim \,\partial _\mu \Lambda ^\mu \left( x\right)
\,\,,  
\label{app-a15}\end{equation}
with $w(x)$ and $\Lambda^\mu(x)$ given by (\ref{a16} - \ref{a17}).

We now come to scale invariance, in the flat limit. 
The classical
dilatation current ${\cal D}^\mu $ can be defined as 
\begin{equation}
{\cal D}^\mu \left( x\right) =x^\lambda \Theta _\lambda ^{~\mu }\left(
x\right) -x^\lambda \left[ w_\lambda ^{~\mu }\left( x\right) \Sigma \right]
-\Lambda ^\mu \left( x\right) \,\,.  \label{dilata}
\end{equation}

It obeys, up to mass terms, dimensionful couplings and total derivatives,
the conservation Ward identity, a direct consequence of (\ref{a1}) and 
(\ref{app-a15}): 
\begin{equation}
\,\partial _\mu {\cal D}^\mu \sim w_{_D}\left( x\right) \Sigma \,\,,
\label{dilata1}
\end{equation}
where 
\begin{equation}
w_{_D}\left( x\right) = w^{\rm trace} - w_\mu ^{~\mu }
=\sum\limits_{{\rm all\,fields}}\left[ \left( d_\Phi
+x\cdot \partial \right) \Phi \right] \frac \delta {\delta \Phi }\,\,,
\label{a17'}
\end{equation}
is the local operator for the scale transformations and $d_\Phi$ is the
dimension of the fields surviving in the flat limit (see Table 1).


\subsection*{Yang-Mills Chern-Simons Theory:}

The equations (\ref{a1} -- \ref{a10}) for the conservation of the
energy-momentum tensor are unchanged -- except the matter terms which are now
absent. The expression for the right-hand side of (\ref{int-theta1}) is
modified into 
\begin{equation}
N_e\Sigma =\left( -\frac 12N_A+\frac 12N_{A^{*}}+\frac 32N_b+\frac 32N_
{\bar{c}}-\frac 12N_c+\frac 12N_{{c}^{*}}\right) \Sigma \,+\left( m\partial
_m+\frac 12g\partial _g\right) \Sigma \,\,.  \label{a18}
\end{equation}
with now a strict equality sign since we keep the contributions from all
terms of the action (\ref{g-inv-action}).

 The correspondent of the local identity \equ{app-a15} reads
\[
w(x)\Sigma = \Lambda(x) \ ,
\]
with $w$ and $\Lambda$ given by (\ref{ym-cl-tr1} -- \ref{lambda})

 Finally, as in (\ref{dilata1}), 
we obtain the broken conservation Ward identity 
\[
\,\partial _\mu {\cal D}^\mu =w_{_D}\left( x\right) \Sigma +\Lambda \left(
x\right) \,\,. 
\]
The local dilatation operator $w_{_D}\left( x\right) $ is given by (\ref
{a17'}), but the dimensions $d_\Phi $ given in Table 2.

\setcounter{equation}{0} \setcounter{section}{2} 

\section*{Appendix B: Definition of the Bilinear Insertion \protect\\
       (\ref{splitting}) Through the Wilson-Zimmermann \protect\\
                      Short Distance Expansion}

The aim is to construct two local insertions $m_A(x)$ and $n_A(x)$, 
of ghost
number $-1$ and 0, respectively, invariant -- or covariant --
under BRS, the diffeomorphisms and the local lorentz transformations, 
and such that 
\begin{equation}
n_A(x)\cdot \Gamma ={\cal B}_\Gamma \left[ m_A(x)\cdot \Gamma \right] \,,
\label{n=bm}
\end{equation}
and 
\begin{equation}
\displaystyle{\int }d^3x\,n_A(x)\cdot \Gamma ={\cal N}_A\Gamma \,,
\label{int-n}
\end{equation}
where ${\cal N}_A$ is the counting operator defined by (\ref{count-op}). The
classical solution of the problem is given by 
\begin{equation}
\begin{array}{l}
m_A(x)={\hat{A}}^{*\mu }A_\mu \,, \\[3mm] 
n_A(x)=A_\mu \left( {\displaystyle{\frac{\delta \Sigma }{\delta A_\mu }}}
+eg^{\mu \nu }\partial _\nu b\right) -{\hat{A}}^{*\mu }{\displaystyle{\frac{
\delta \Sigma }{\delta {\hat{A}}^{*\mu }}}}\,,
\end{array}
\label{class-m-n}
\end{equation}
with ${\hat{A}}^{*\mu }$ defined by (\ref{combination1}).

In order to find their quantum extensions, let us consider the point
splitted insertions introduced in (\ref{splitting}): 
\begin{equation}
\begin{array}{l}
m_A^e(x)\cdot \Gamma ={\hat{A}}^{*\mu }(x_{+})A_\mu (x_{-})\cdot \Gamma \,,
\\[3mm] 
n_A^e(x)\cdot \Gamma ={\cal B}_\Gamma \left[ m_A^e(x)\cdot \Gamma \right] \,,
\end{array}
\label{split-m-n}
\end{equation}
with $x_{\pm }^\mu =x^\mu \pm \epsilon ^\mu$, and let us show that the
limit $\epsilon \to 0$ exists and solves the problem.

The Wilson-Zimmermann short distance expansion~\cite{wil-zim} yields%
\footnote{%
The short distance expansion has been rigorously proved only for theories
without massless fields. The extension to massless fields is assumed 
to hold.} 
\begin{equation}
m_A^e(x)\cdot \Gamma =\displaystyle{\sum_K^{}}f_K(\epsilon )B_K(x)\cdot
\Gamma +R^e(\epsilon ,x)\,,  \label{wz-exp}
\end{equation}
where $R^e$ goes to zero in the limit $\epsilon \to 0$. The $B_K(x)$'s are
all possible renormalized insertions (Zimmermann's normal 
products~\cite{Zimmerman}) of dimension $\le $ 3 
(the dimension of $m_A$), and the
coefficients $f_K(\epsilon )$ have a short distance behaviour 
\[
f_K(\epsilon )\sim |\epsilon |^{-d_K}\quad \mbox{up to logarithmic factors}
\,,\quad d_K\le 3-\mbox{dim}(B_K)\,. 
\]
Taking into account all the constraints which must be obeyed by $m_A$, we
see that the list of possible operators $B_K$ reduces to quantum extensions
of the three following expressions: 
\begin{equation}
\begin{array}{ll}
B_1={{\hat{A}}^{*\mu }}A_\mu & d_1\le 0\,, \\[3mm] 
B_2=\bar{\Psi}{}^{*}\Psi +\bar{\Psi}\Psi ^{*}\qquad & d_2\le 0\,, \\[3mm] 
B_3=\varphi ^{*}\varphi & d_3\le 0\,.
\end{array}
\label{wz-basis}
\end{equation}
Now, we observe that the integral of $m_A^e$ can be written, after partial
integration of the term involving the antighost field $\bar{c}$, and use of
the gauge condition (\ref{landau1}), as 
\[
\displaystyle{\int }d^3x\,m_A^e(x)\cdot \Gamma =\displaystyle{\int }
d^3x\left( A^{*\mu }(x_{+})A_\mu (x_{-})-\bar{c}(x_{+}){\displaystyle
{\ \frac{\delta \Gamma }{\delta b(x_{-})}}}\right) \,, 
\]
which has the well defined limit 
\begin{equation}
\displaystyle{\int }d^3x\left( A^{*\mu }(x)A_\mu (x)-\bar{c}(x)
{\displaystyle {\frac{\delta \Gamma }{\delta b(x)}}}\right) \,,
\label{lim-int-m}
\end{equation}
for $\epsilon \to 0$. Since there is no total derivative term in the
expansion (\ref{wz-exp}), we conclude from this that all three coefficients 
$f_K(\epsilon )$ are regular at $\epsilon =0$. This proves the existence of
the limit 
\begin{equation}
m_A(x)\cdot \Gamma =\lim_{\epsilon \to 0}m_A^e(x)\cdot \Gamma \,,
\label{lim-m}
\end{equation}
with the integral $\int d^3x\,m_A(x)\cdot \Gamma $ being equal to the
expression (\ref{lim-int-m}).

The existence of the limit 
\begin{equation}
n_A(x)\cdot \Gamma =\lim_{\epsilon \to 0}n_A^e(x)\cdot \Gamma \,,
\label{lim-n}
\end{equation}
with the properties (\ref{n=bm}) and (\ref{int-n}) follows by acting with
the BRS operator ${\cal B}_\Gamma $ on (\ref{lim-m}).


\end{document}